\documentclass[prl,aps,twocolumn,showpacs,preprintnumbers,amsmath,amssymb]{revtex4}
\usepackage{epsfig}
\usepackage{color}
\usepackage{ifthen}
\usepackage{xfrac}
\usepackage{graphicx}
\usepackage{amsmath}
\usepackage{amssymb}

\usepackage{graphicx}% Include figure files
\usepackage{dcolumn}% Align table columns on decimal point
\usepackage{bm}% bold math
\def\Dsl{\hbox{/\kern-.6700em\it D}} % D slash
\def\dsl{\hbox{/\kern-.5300em$\partial$}}

\def\eqa{\begin{eqnarray}}
\def\eeqa{\end{eqnarray}}
\def\eq{\begin{equation}}
\def\eeq{\end{equation}}
\def\be{\begin{equation}}
\def\ee{\end{equation}}
\def\bea{\begin{eqnarray}}
\def\eea{\end{eqnarray}}

\newcommand{\dslash}{\not{\hbox{\kern-2pt $\partial$}}}
\newcommand{\pslash}{\not{\hbox{\kern-2.3pt $p$}}}
 \newtoks\nslashfraction
 \nslashfraction={.13}
 \newcommand{\nslash}[1]{\setbox0\hbox{$ #1 $}
   \setbox0\hbox to \the\nslashfraction\wd0{\hss \box0}/\box0 }

\def\lsim{\mathrel{\rlap{\lower4pt\hbox{\hskip1pt$\sim$}}
    \raise1pt\hbox{$<$}}}                % less than or approx. symbol
\def\gsim{\mathrel{\rlap{\lower4pt\hbox{\hskip1pt$\sim$}}
    \raise1pt\hbox{$>$}}}   

\begin{document}

\preprint{}

\title{A Consistency Relation for Single-Field Inflation \\with Power Spectrum Oscillations}
\author{Mark G. Jackson$^{1,2,3}$ and Gary Shiu$^{4,5}$}

\affiliation{$^1$Institut d'Astrophysique de Paris, UMR CNRS 7095,}
\affiliation{Universit\'{e} Pierre et Marie Curie, 98bis boulevard Arago, 75014 Paris, France}
\affiliation{$^2$Department of Physics, University of Illinois at Urbana-Champaign, Urbana, IL 61801}
\affiliation{$^3$Paris Centre for Cosmological Physics and Laboratoire AstroParticule et Cosmologie, \\Universit\'{e} Paris 7-Denis Diderot, Paris, France}
\affiliation{$^4$Department of Physics, University of Wisconsin-Madison, Madison, WI 53706}
\affiliation{$^5$Department of Physics and Institute for Advanced Study, \\Hong Kong University of Science and Technology, Hong Kong}

\date{\today}


\begin{abstract}
\noindent
We derive a theoretical upper bound on the oscillation frequency in the scalar perturbation power spectrum of single-field inflation. Oscillations are most naturally produced by modified vacua with varying phase. When this phase changes rapidly, it induces strong interactions between the scalar fluctuations.  If the interactions are sufficiently strong the theory cannot be evaluated using perturbation theory, hence imposing a limit on the oscillation frequency.  This complements the bound found by Weinberg governing the validity of effective field theory.  The generalized consistency relation also allows one to use squeezed configurations of higher-point correlations to place constraints on the power spectrum oscillations.
\end{abstract}

\maketitle

%%%%%%%%%%%%%%%%%%%%%%%%
\subsection{Introduction}
%%%%%%%%%%%%%%%%%%%%%%%%
The scalar fluctuation $\zeta(t,{\bf x})$ is the gauge-invariant perturbation \cite{Bardeen:1983qw,Salopek:1990jq} which appears in the nearly FRW metric as
\begin{equation}
\label{background}
 ds^2 = -dt^2 + a(t)^2 e^{2 \zeta} d{\bf x}^2
 \end{equation}
once gauge-fixing has been imposed.  Much attention has been given to the late-time correlation functions of $\zeta$ in inflating backgrounds.  For example, the two-point correlation measures the power spectrum $P_\zeta(k)$, 
\begin{equation}
\label{twopoint}
 \langle \zeta_{{\bf k}_1} \zeta_{{\bf k}_2} \rangle_0 = P_\zeta(k_1) (2 \pi)^3 \delta^3({\bf k}_1+{\bf k}_2). 
 \end{equation}
The subscript indicates that this is evaluated in the background with the classical scalar field set to zero, $\zeta^B=0$. It has been experimentally determined \cite{Hinshaw:2012fq} that the power spectrum is very nearly scale-invariant and is well-described by the following form,
\begin{equation}
\label{plainansatz}
P_\zeta(k) \approx \left( \frac{H^2}{ {\dot \phi}} \right)^2 \frac{1}{2k^3}  \left( \frac{k}{k_0} \right)^{n_s-1}
\end{equation}
where $H \equiv {\dot a}/a$ is the energy scale of inflation, $\phi$ is the background field responsible for inflation, $n_s \approx 0.96$ is the `tilt' of scale-dependence, and $k_0$ is a `pivot point' which should be taken somewhere near the middle of observationally accessible $k$.  The amplitude of the power spectrum has been measured to be roughly 
\[ P_\zeta \sim 10^{-10}. \]
The bispectrum, or three-point correlation, probes self-interactions of the field.  One such type of interaction is that which is local in position space \cite{Komatsu:2001rj},
\[ \zeta(x) \equiv \zeta_G (x) + \frac{3}{5} f^{\rm loc}_{\rm NL}  \zeta^2_G (x) \]
 where $\zeta_G$ is a Gaussian-distributed field and $f^{\rm loc}_{\rm NL}$ is a constant parameterizing this local non-linearity.  Fourier transforming this gives the following relation,
\begin{eqnarray}
\label{bilocal}
 \langle \zeta_{{\bf k}_1} \zeta_{{\bf k}_2} \zeta_{{\bf k}_3} \rangle &=& \frac{6}{5} f_{\rm NL}^{\rm loc} \left[ P_\zeta(k_1) P_\zeta(k_3) \right. \\
 \nonumber
&& \hspace{-0.4in} \left. + P_\zeta(k_2) P_\zeta(k_1) + P_\zeta(k_3) P_\zeta(k_2) \right].
 \end{eqnarray}
Here, and henceforth, we will omit the factors of  $ (2 \pi)^3  \delta^3(\sum_i {\bf k}_i)$.
 
A single-field consistency relation was noted by Maldacena \cite{Maldacena:2002vr} as follows.  If one of the momentum magnitudes, say $k_3$, is much smaller than the others then that mode will freeze out into a background configuration in the spatial metric much earlier. The bispectrum can then be evaluated with $\zeta_{{\bf k}_3}$ simply rescaling the magnitudes of the other momentum.

A formal proof of this was provided in \cite{Cheung:2007sv}.  The two-point correlation (\ref{twopoint}) in a background (\ref{background}) with $|\zeta|~\ll~1$ will then be close to the homogeneous FRW case,
\begin{eqnarray}
\label{fieldexpansion}
\langle \zeta( {\bf x}_1 )  \zeta( {\bf x}_2) \rangle_{\zeta^B} &\approx& \langle \zeta( {\bf x}_1 )  \zeta( {\bf x}_2) \rangle_0 \\
\nonumber
&+& \zeta^B \left. \frac{\delta}{\delta \zeta^B} \right|_0  \langle \zeta( {\bf x}_1 ) \zeta( {\bf x}_2) \rangle_{\zeta^B} + \cdots .
\end{eqnarray}
If we further assume that the short-distance difference ${\bf x}_S~\equiv~{\bf x}_1 - {\bf x}_2$ is much smaller than the long-distance average position ${\bf x}_L~\equiv~({\bf x}_1 + {\bf x}_2)/2$ then we can approximate the background value by the average position, $\zeta^B( {\bf x}) \approx \zeta^B( {\bf x}_L)$.  Using the fact that $\zeta$ represents a spatial rescaling, this makes the above expansion
\begin{eqnarray*}
\langle \zeta( {\bf x}_1 )  \zeta( {\bf x}_2) \rangle_{\zeta^B} &\approx& \langle \zeta( {\bf x}_S/2  )  \zeta( -{\bf x}_S/2) \rangle_0 \\
&& \hspace{-0.7in} + \zeta^B( {\bf x}_L) \frac{d}{d \ln x_S}  \langle \zeta( {\bf x}_S/2  )  \zeta( -{\bf x}_S/2) \rangle_0 + \cdots .
\end{eqnarray*}
Now Fourier transforming with respect to ${\bf x}_1$ and ${\bf x}_2$, defining ${\bf k}_S~\equiv~{\bf k}_1 - {\bf k}_2$ and ${\bf k}_L~\equiv~({\bf k}_1 + {\bf k}_2)/2$ and performing some algebra gives the relation
\begin{equation}
\label{kcondition}
\langle \zeta_{{\bf k}_1}  \zeta_{{\bf k}_2}\rangle_{\zeta^B} \approx \langle \zeta_{{\bf k}_S}  \zeta_{-{\bf k}_S} \rangle_0 -\frac{ \zeta_{{\bf k}_L} }{k_S^3} \frac{d}{d \ln k_S} \left(k_S^3  \langle \zeta_{{\bf k}_S} \zeta_{-{\bf k}_S} \rangle_0 \right) + \cdots
\end{equation}

Multiplying both sides by $\zeta_{{\bf k}_3}$ and taking the correlation gives the relation
\begin{equation}
\label{consistency}
 \langle \zeta_{{\bf k}_1} \zeta_{{\bf k}_2} \zeta_{{\bf k}_3} \rangle_{k_3 \ll k_1,k_2} \approx -P_\zeta(k_L) P_\zeta(k_S) \frac{d\ln [ k_S^3 P_\zeta(k_S) ]}{d \ln k_S}.
 \end{equation}
Assuming the ansatz (\ref{plainansatz}), this makes
\[  \langle \zeta_{{\bf k}_1} \zeta_{{\bf k}_2} \zeta_{{\bf k}_3} \rangle_{k_3 \ll k_1,k_2} \approx -(n_s-1) P_\zeta(k_L) P_\zeta(k_S) . \]
Comparing this to (\ref{bilocal}) with $n_s \sim \mathcal O(1)$ means a measurement of $f_{\rm NL}^{\rm local} \gsim 1$ eliminates most (but apparently not all \cite{Chen:2013aj}) models of single-field inflation \cite{Creminelli:2004yq}.  Note that there has been no assumption of the field dynamics (such as slow-roll), this is a direct consequence of a single field being the only ``clock" in the system.  This powerful tool to relate the bispectrum to the spectrum can also be derived from a dual conformal field theory perspective \cite{Schalm:2012pi} and has been applied in various examples \cite{Chen:2006nt,Ganc:2010ff, RenauxPetel:2010ty}. 

Now suppose that the power spectrum \emph{did not} fit the form (\ref{plainansatz}) but instead contained oscillatory features, perhaps of the form
\[ P_\zeta \sim \frac{P_0}{k^3} \left[ 1 + \beta \cos \left( \omega \ln \frac{k}{k_0} \right) \right]. \]
Then the logarithmic derivative is
\begin{equation}
\label{bigosc}
 \frac{d \ln (k^3 P_\zeta)}{d \ln k} \sim \beta \omega \sin \left( \omega \ln \frac{k}{k_0} \right).
 \end{equation}
By tuning $\beta$ and $\omega$ one could produce arbitrarily small oscillations in the power spectrum yet arbitrarily large non-Gaussianity, and so could easily violate the ``spirit" of the usual consistency condition.  Of course this also means the non-Gaussianity will be highly oscillatory and therefore not truly of the ``local" shape, making it more difficult to constrain \cite{Flauger:2010ja}, but this is a practical issue not a theoretical one.
While there have been previous studies of oscillations in the bi- and trispectrum they tended to focus either on the enfolded shape \cite{Holman:2007na,Meerburg:2009ys,Meerburg:2009fi,Meerburg:2010ca} or of theoretically-motived origin \cite{Jackson:2012fu}.

Such oscillations are most often the result of vacuum choice (or equivalently, boundary conditions) which produce an interference pattern between the positive and negative frequency modes.  Recall that the scalar field is quantized as
\[ \zeta_{\bf k}(\tau) = \frac{H^2}{|{\dot \phi}|  \sqrt{2k^3} } \left[ a^\dagger_{\bf k} \left( 1+ i k \tau \right) e^{-ik \tau} +  a_{\bf k} \left( 1- i k \tau \right) e^{ik \tau} \right] \]
where we have introduced the conformal time 
\[ a(\tau) d \tau \equiv dt. \]  
A class of vacuum states are parameterized by complex $\alpha_{\bf k}, \beta_{\bf k}$ such that 
\[ \left( \alpha_{\bf k} a_{\bf k} + \beta_{\bf k} a^\dagger_{\bf k} \right) | \alpha_{\bf k}, \beta_{\bf k} \rangle = 0. \]
Normalization requires $ \left| \alpha_{\bf k} \right|^2 - \left| \beta_{\bf k} \right|^2 = 1$.  The special choice $\alpha_{\bf k} = 1, \beta_{\bf k}=0$ is known as the Bunch-Davies vacuum \cite{Bunch:1978yq} and has been assumed for most previous analyses of the consistency relation (\ref{consistency}).  The squeezed limit of single-field inflation with modified vacuum states has been studied in \cite{Agullo:2010ws, Ganc:2011dy, Gong:2013yvl,Aravind:2013lra} and obtained various bounds on $|\beta_{\bf k}|$ based on the backreaction of excited states.  However, these largely assumed slowly-varying vacua as in (\ref{bigosc}).  In \cite{Chen:2008wn, Chen:2010bka} rapid oscillations were obtained from resonance interactions but which could be effectively described by a vacuum rotation \cite{Flauger:2010ja, Flauger:2013hra}.

In this article we focus attention on the fact that such rapid oscillations could overwhelm the prefactor $n_s-1$, leading to an apparent violation of the consistency relation, but is simply the result of including the modified vacuum.  This also produces non-trivial relations between the various correlation functions which could be used to establish bounds on the oscillation parameters.

%%%%%%%%%%%%%%%%%%%%%%%%
\subsection{NPH Oscillations}
%%%%%%%%%%%%%%%%%%%%%%%%
Consider the `Minimal Uncertainty State' \cite{Easther:2002xe} parameterized by
\begin{equation}
\label{beta}
\beta_{\bf k} = \frac{\beta}{2} e^{i \frac{M}{H} \ln \frac{k}{k_0} }. 
\end{equation}
We use $H/M$ to denote the wavenumber since this type of vacuum typically arises from `new' physics arising at energy scale $M$, referred to as the New Physics Hypersurface (NPH) \cite{Greene:2005aj,arXiv:1007.0185,arXiv:1104.0887,Jackson:2012qp}.  We further assume that
\begin{equation}
\label{oscvalues}
 H/M \ll 1, \hspace{0.5in} \beta \ll 1 . 
 \end{equation}
This modifies the power spectrum to include oscillations of magnitude $\beta$ and periodicity in $\ln k$ with wavelength $H/M$:
\[ P_\zeta(k) \approx  \left( \frac{H^2}{ {\dot \phi}} \right)^2 \frac{1}{2k^3} \left( \frac{k}{k_0} \right)^{n_s-1} \left[ 1 + \beta \cos \left( \frac{M}{H} \ln \frac{k}{k_0} \right) \right] . \]
Such oscillations have been searched for using \emph{WMAP} data but not yet found \cite{Martin:2004yi, Meerburg:2011gd}.  The \emph{Planck} data \cite{planck} will be much more precise and so may allow detection.

Here we will consider a particular limit of such states, one in which the oscillations are so rapid that they dominate the scale-dependence of $P_\varphi(k)$, 
\[\frac{ \beta M }{H} \gg n_s-1 . \]
In this limit the bispectrum relation (\ref{consistency}) becomes
\begin{equation}
\label{bi}
\hspace{-0.1in} \langle \zeta_{{\bf k}_1} \zeta_{{\bf k}_2} \zeta_{{\bf k}_3} \rangle_{k_3 \ll k_1,k_2} \approx \frac{\beta M}{H} P_\zeta(k_1) P_\zeta(k_3) \sin \left( \frac{M}{H} \ln \frac{k_1}{k_0} \right) .
 \end{equation}
If we define the slow-roll parameter $\epsilon \equiv {\dot \phi}^2/ 2H^2 M_{\rm pl}^2 \sim H^2 / P_\zeta M^2_{\rm pl}$, it can be seen that $n_s-1 = -4 \epsilon$, whereas $\beta M/H$ is clearly independent of this parameter.

The fact that (\ref{bi}) is not slow-roll suppressed may appear paradoxical, since it was shown \cite{Maldacena:2002vr} that all single-field interactions could only produce a local bispectrum which is $\mathcal O(\epsilon)$.  The reason for this can be seen as follows.  The choice of vacuum is implemented via a boundary term \cite{Schalm:2004qk} of the primordial field fluctuation $\varphi$ which is schematically of the form
\[ S_{\rm b.c.} \sim \int_{\eta_k} d^3 {\bf k} \ \beta_{\bf k} \left| \varphi_{\bf k} \right|^2 \]
where $\eta_k = - M/Hk$ is the time at which the mode has physical energy $M$ and
\[ \beta_{\bf k} = \int \frac{d^3 {\bf x}}{(2 \pi)^3} \beta( {\bf x}) e^{-i x^i k^j g_{ij}} . \]
The primordial fluctuation is converted to the gauge-invariant quantity as $\varphi \sim \sqrt{\epsilon} \zeta$.  Once $g_{ij}$ is expanded via (\ref{background}), additional factors of $\zeta$ will appear but without any slow-roll suppression, leading to an action of the form
\begin{equation}
\label{betaexpansion}
 S_{\rm b.c.} \sim \int_{\eta_k} d^3 {\bf k} \ \beta_{{\bf k}_0} \left[ 1 - \zeta \left. \frac{d \ln \beta_{\bf k}}{d \ln k} \right|_{{\bf k}_0} + \cdots \right] \epsilon \zeta^2.
 \end{equation}
Substituting (\ref{beta}) into this expression immediately recovers the bispectrum (\ref{bi}).  This $\mathcal O(\epsilon)$ cubic interaction is one $\epsilon$-power less than the $\epsilon^2 \zeta^3$ interactions appearing in the bulk.  This is why (\ref{bi}) is not slow-roll suppressed.

It is important to note that applying the consistency relation to modified boundary conditions requires not only that $\zeta_{{\bf k}_3}$ freezes out before $\zeta_{{\bf k}_{1,2}}$ do, it is also necessary that it freezes out before these boundary conditions for $\zeta_{{\bf k}_{1,2}}$ are defined. This means that
\[ k_3 \eta_{1,2} = \frac{k_3 M}{ k_{1,2} H} \ll 1. \]
This is a factor $M/H$ stronger than the usual condition $k_3 \ll k_{1,2}$, and was also noted for the resonance model in \cite{Flauger:2013hra}.  If this more restrictive condition is not obeyed, the boundary term will not be included in the spatial rescaling due to $\zeta_{{\bf k}_3}$ and hence will not be included in the estimation of the bispectrum.  The fact that \emph{Planck} is sensitive to momentum scales over a wider range than its predecessor $(l \sim 2500)$ is essential in testing this very squeezed limit relation.

Performing the same procedure for the trispectrum yields an even more dramatic result.  Continuing the expansion of (\ref{kcondition}) to second order gives
\[ \langle \zeta_{{\bf k}_1}  \zeta_{{\bf k}_2}\rangle^{(2)}_{\zeta^B} =  \frac{ \zeta_{{\bf k}_L-{\bf q}} \zeta_{{\bf q}} }{2k_S^3} \frac{d^2}{d (\ln k_S)^2} \left(k_S^3  \langle \zeta_{{\bf k}_S} \zeta_{-{\bf k}_S} \rangle_0 \right). \] 

Taking the 4-point analogue of the squeezed limit means evaluating the power spectrum with two modes frozen into their background values \cite{Huang:2006eha}, and so the trispectrum is dominated by the second derivative of the oscillations:
\begin{eqnarray*}
\label{tri}
&& \hspace{-0.3in} \langle \zeta_{{\bf k}_1} \zeta_{{\bf k}_2} \zeta_{{\bf k}_3} \zeta_{{\bf k}_4} \rangle_{k_3,k_3 \ll k_1,k_2} \\
 &\approx&- \beta \left( \frac{M}{H} \right)^2  \frac{P_\zeta(k_1) P_\zeta(k_3) P_\zeta(k_4)}{2} \ \cos \left( \frac{M}{H} \ln \frac{k_1}{k_0} \right) . 
\end{eqnarray*}
Of course this immediately generalizes to the $n$-point correlation (ignoring the sign and not distinguishing between sine and cosine),
\begin{eqnarray*}
\label{npoint}
&& \hspace{-0.3in} \langle \zeta_{{\bf k}_1} \zeta_{{\bf k}_2} \zeta_{{\bf k}_3} \cdots \zeta_{{\bf k}_n} \rangle_{k_3, \ldots, k_n \ll k_1,k_2} \\
 && \hspace{-0.4in} \approx \beta \left( \frac{M}{H} \right)^{n-2}  \frac{P_\zeta(k_1)}{(n-2)!}  \cos \left( \frac{M}{H} \ln \frac{k_1}{k_0} \right) \prod_{i=3}^n P_\zeta(k_i)  . 
\end{eqnarray*}

%%%%%%%%%%%%%%%%%%%%%%%%
\paragraph{Theoretical Constraints}
%%%%%%%%%%%%%%%%%%%%%%%%

We saw in (\ref{betaexpansion}) that if the boundary conditions change with position, this produces higher-order interactions in $\zeta$.  These are then probed using the free field contraction $\langle \zeta^2 \rangle \sim P_\zeta$.  If the spatial variation is too quick, meaning the interactions are too strong, this cannot be evaluated perturbatively.  For $H/M \lsim P_\zeta$ the $n$-point correlation becomes larger than the $(n-1)$-point, fixing the bound
\[  M  \lsim \frac{H}{P_\zeta} \sim 10^{10} H . \]
This is to be compared to Weinberg's bound on effective field theory \cite{Weinberg:2008hq}, $M \gsim \sqrt{\epsilon} M_{\rm pl}$.  If the oscillations are induced by integrating out heavy physics at scale $M$, this fixes a lower limit $M \gsim H/ \sqrt{P_\zeta}$.  This bounds the frequency as
\[ 10^5 \lsim \frac{M}{H} \lsim 10^{10} . \]
It is interesting that this bound does not contain the magnitude $\beta$, merely the frequency.  This is because it is due only to the \emph{relative} change of the field as a function of scale.  The Bunch-Davies case $\beta_{\bf k} = 0$ of course also satisfies the constraint.

%%%%%%%%%%%%%%%%%%%%%%%%
\paragraph{Observational Constraints}
%%%%%%%%%%%%%%%%%%%%%%%%

The close relationship between the power spectrum and higher-point correlations also places strong observational constraints on oscillations.  One could use the power spectrum to place a bound on both the magnitude of the oscillations to a level of $\beta \lsim 10^{-2}$.  It was estimated in \cite{Greene:2005aj} that upcoming data could in principle resolve oscillation frequencies as high as $M/H \sim 10^2$.  The bispectrum condition (\ref{bi}) implies that a precision measurement will place a bound on $\beta M/H$.  A sensitive enough measurement would exclude a significant portion of $\beta-M/H$ parameter space.  Similarly, measurement of the trispectrum would place an analogous bound on $\beta (M/H)^2$, excluding even more parameter space.  Subsequent measurements of higher-point correlations would continue this, the $n$-point correlation asymptotically producing a bound simply on $M/H$ as $n \rightarrow \infty$.  The combined bounds are shown in Figure 1.  This figure is meant to be merely schematic, as the experimental bounds on the bi- and trispectrum are not fully known.

\begin{figure}
\begin{center}
\includegraphics[width=2.5in]{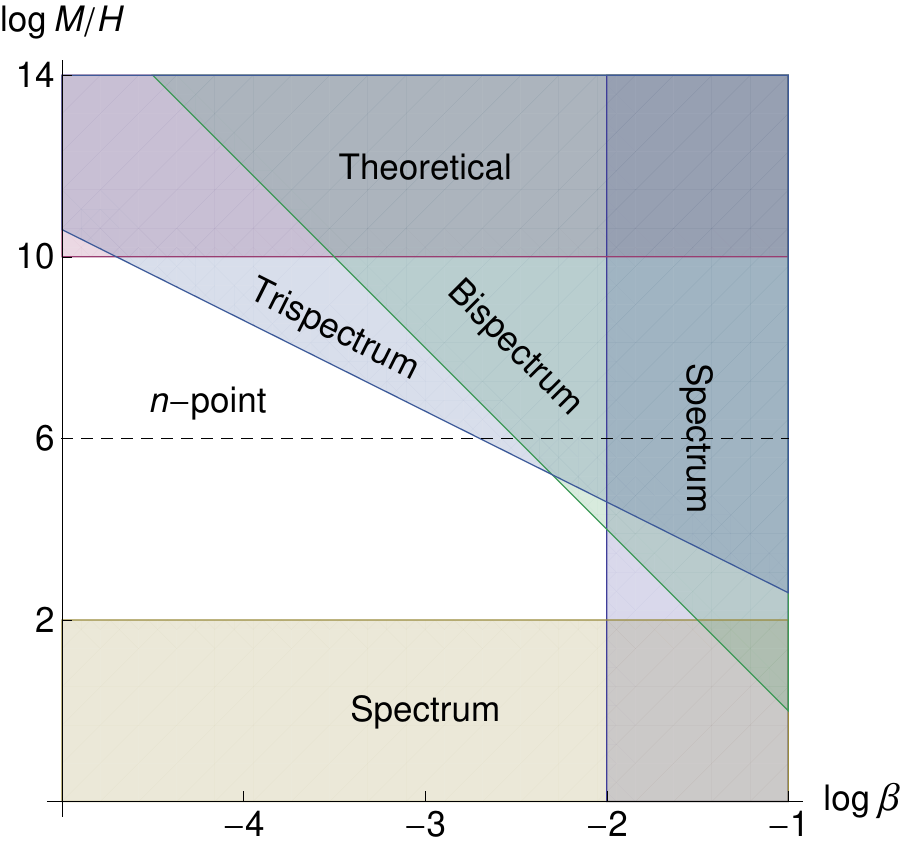}
\caption{NPH parameter space bounds.  Analogous bounds exist for BEFT parameter space by replacing $M/H \rightarrow \mathcal C k_{\rm max}$ and $ \beta \rightarrow\mathcal A k_{\rm max}$.}
\end{center}
\end{figure}

%%%%%%%%%%%%%%%%%%%%%%%%
\subsection{BEFT Oscillations}
%%%%%%%%%%%%%%%%%%%%%%%%
Boundary Effective Field Theory (BEFT) models new physics which occurs at one moment in time, such as a discontinuity in single-field inflation \cite{Starobinsky:1992ts} or a sharp turn in multi-field inflation \cite{Shiu:2011qw}.  These generically produce oscillations which have magnitude and periodicity which are linear in $k$.  Even though we are (by definition) only considering single-field inflation models we may still consider BEFT theories and their signatures.  

The power spectrum will now have the form
\[ P_\zeta(k) \approx \left( \frac{H^2}{{\dot \phi}} \right)^2  \frac{1}{2k^3} \left( \frac{k}{k_0} \right)^{n_s-1} \left[ 1 + \mathcal A k \cos \left( \mathcal C k \right) \right] . \]
Since all boundary conditions are defined at nearly the same time $\eta_{\rm BEFT}$, the condition to ensure that $\zeta_{{\bf k}_3}$ has frozen out into a background condition is strengthened to 
\[ k_3 \eta_{\rm BEFT} \ll 1. \]
For sufficiently rapid oscillations the squeezed bispectrum is now
\[   \langle \zeta_{{\bf k}_1} \zeta_{{\bf k}_2} \zeta_{{\bf k}_3} \rangle_{k_3 \ll k_1,k_2} \approx \mathcal A \mathcal C k_1^2 P_\zeta(k_1) P_\zeta(k_3) \sin \left( \mathcal C k_1 \right) . \]
Subsequent correlations have a magnitude containing additional factors of $\mathcal C k_1 P_\zeta $.

%%%%%%%%%%%%%%%%%%%%%%%%
\paragraph{Theoretical Constraints}
%%%%%%%%%%%%%%%%%%%%%%%%
The effective oscillation magnitude is $\mathcal A k$ and effective frequency $\mathcal C k$.  The theoretical bound in this case then requires that $\mathcal C k_{\rm max}~\lsim~P_\zeta^{-1}$, where $k_{\rm max}$ is the maximally measurable wavenumber, implying $\mathcal C k_{\rm max} \lsim 10^{10}.$

%%%%%%%%%%%%%%%%%%%%%%%%
\paragraph{Observational Constraints}
%%%%%%%%%%%%%%%%%%%%%%%%
The identical procedure can be used to obtain the higher-point correlation functions for BEFT as used for NPH, using the above parameter replacements.  This implies that an identical bounding of parameter space can be made by making the replacements  $M/H \rightarrow \mathcal C k_{\rm max}$ and $ \beta \rightarrow\mathcal A k_{\rm max}$.

%%%%%%%%%%%%%%%%%%%%%%%%
\subsection{Conclusion}
%%%%%%%%%%%%%%%%%%%%%%%%
We generalized the single-field consistency condition to include rapid oscillations as a result of modified boundary conditions.  This produces an upper bound on the oscillation frequency as well as non-trivial relations between the power spectrum and higher-point correlations.  

A precise measurement of the local-shape bispectrum with oscillations is then exceptionally important.  While upcoming data from \emph{Planck} should be sensitive to $\Delta f^{\rm local}_{\rm NL} \sim 6$, in practice this is not of much use because it would be largely orthogonal to a local shape containing oscillations \cite{Flauger:2010ja}.  This strongly motivates developing efficient search technique for correlation functions containing oscillations such as in \cite{Jackson:2013mka}.

%%%%%%%%%%%%%%%%%%%%%%%%
{\bf Acknowledgments:}
%%%%%%%%%%%%%%%%%%%%%%%%
We thank X. Chen, E. Komatsu, K. Schalm, D. Steer, H. Tye, and B. Wandelt for useful discussions.  MGJ would like to thank the Hong Kong University of Science and Technology for hospitality during which this work was completed.  The work of GS was supported in part by DOE grant DE-FG-02-95ER40896.

\end{document}